\documentclass[aps,prb,floats,epsfig,twocolumn]{revtex4-2}

\usepackage{graphicx}
\usepackage{amsmath}
\usepackage{amssymb}
\usepackage{braket}
\usepackage{bbold}
\usepackage{hyperref}
\usepackage{subfigure}
\usepackage[usenames,dvipsnames]{color}
\usepackage{graphicx}
\usepackage{subfigure}
\usepackage[mathscr]{euscript}

\newcommand{\Tr}{\rm{Tr}}

\newcommand{\tb}{\textcolor{black}}

\begin{document}

\title{Disconnected entanglement entropy as a marker of edge modes in a 
periodically driven Kitaev chain}
\author{Saikat Mondal$^1$}
\email{msaikat@iitk.ac.in}
\author{Diptiman Sen$^2$}
\email{diptiman@iisc.ac.in}
\author{Amit Dutta$^1$}
\email{dutta@iitk.ac.in}
\affiliation{$^1$Department of Physics, Indian Institute of Technology,
Kanpur 208016, India \\$^2$Centre for High Energy Physics and Department of Physics, 
Indian Institute of Science, Bengaluru 560012, India}

\begin{abstract}
We study the disconnected entanglement entropy (DEE) of a Kitaev chain in which the
chemical potential is periodically modulated with $\delta$-function pulses within the framework of Floquet theory. For this driving protocol, the DEE of a sufficiently 
large system with open boundary conditions turns out to be integer-quantized, with the 
integer being equal to the number of Majorana edge modes localized at each edge of the chain generated by the periodic 
driving, thereby establishing the DEE as a marker for detecting Floquet Majorana edge modes. Analysing the DEE, we further show that these Majorana edge modes are robust against weak spatial disorder and temporal noise. Interestingly, we find that the DEE may, in 
some cases, also detect the anomalous edge modes which can be generated by periodic
driving of the nearest-neighbor hopping, even though such modes have no topological 
significance and not robust against spatial disorder. We also probe the behaviour of
the DEE for a kicked Ising chain in the presence of an integrability breaking interaction which has been
experimentally realized.
\end{abstract}

\maketitle

\section{Introduction}
\label{sec_intro}

There is a recent upsurge in studies of topological phases of matter~\cite{hasan10, 
qi11, lukasz11, chen13, senthil15, chiu16}. These phases are robust against weak
perturbations due to the existence of a bulk gap, which does not vanish unless the
system crosses a gapless quantum critical point (QCP). In addition, a topological 
phase is characterized by a topological invariant, which remains constant under 
continuous variations of parameters as long as the system remains in the same phase 
and becomes ill-defined at QCPs which separate different phases.

In this regard, the Kitaev chain of spinless fermions (a $p$-wave superconducting 
system in one dimension) is a paradigmatic model that hosts symmetry protected 
topologically non-trivial and 
topologically trivial phases separated by a QCP~\cite{Kitaev01, laumann09}. The 
topological properties of a Kitaev chain with periodic boundary conditions are 
characterized by a topological invariant known as the winding number. The winding 
number assumes non-zero integer-quantized values in the topologically non-trivial 
phase and vanishes in the topologically trivial phase. For a system with open boundary
conditions, the topologically non-trivial phase of the model is manifested in the 
existence of zero-energy Majorana modes localized at the edges; on the contrary, the 
topologically trivial phase does not host Majorana edge modes. The exact solvability 
of the model has been extensively exploited to understand its equilibrium as well as
out-of-equilibrium properties~\cite{Kitaev01, laumann09, DeGottardi11, 
DeGottardi13, dsen13, trs_sen13, rajak14, dutta15, saha17, souvik_open20, souvik2021}.

Concerning the non-equilibrium dynamics of closed quantum systems, periodically 
driven systems have been explored both in the context of 
thermalization~\cite{russomanno12,tomka12, nag14, bukov15, sayak15, nag15, arnabsen16, mukherjee16, eckardt17, dutta2018, takashi19, sen_2021, saikat22} and emergent topology ~\cite{kitagawa10, jiang11, gu11, trif12, dsen13, liu13, DeGottardi13, kundu13, wu13, tong13, andres13, lindner13, katan13, thakurathi14, asboth14, reichl14, perez15, adhip16, roy16, dalla_torre16, russomanno17, yao17, thakurathi17, saha17, molignini17, vega18, molignini18, tilen19, utso_2019, molignini19, molignini20, molignini21, souvik2021} (for recent review 
articles, see Refs.~\cite{bukov15, eckardt17, sen_2021, souvik2021}). For a 
periodically driven Kitaev chain with open boundary conditions, it has been shown that Majorana 
edge modes (zero and $\pi$-modes) can be dynamically generated~\cite{dsen13, saha17, souvik2021} even 
though the instantaneous Hamiltonian 
may remain topologically trivial at all times. In fact, it is the effective Floquet 
Hamiltonian~\cite{floquet83} that determines the non-trivial topology (i.e., the 
existence of Majorana edge modes) of a driven chain. It has been 
observed that the 
number of such dynamical edge modes increases as the drive frequency is reduced. 
Further, a strong (having a sufficiently large amplitude) periodic modulation of the 
hopping parameter can produce some ``anomalous" edge modes with non-zero Floquet 
quasienergies~\cite{saha17}. However, the anomalous edge modes are not Floquet Majorana modes. These modes have no topological 
significance and topological invariants like the winding number miss them completely. Furthermore, as we elaborate in this work, these modes are not robust against
weak spatial disorder.

There have been several attempts to characterize the out-of-equilibrium topology of one-dimensional systems (for instance, the Kitaev chain and the Su-Schrieffer-Heeger chain) using the dynamical winding number calculated from the instantaneous wave function of the system~\cite{cooper18, souvik19, souvik2019}. In a periodically driven Kitaev chain, the corresponding winding number~\cite{dsen13} is calculated from the Floquet Hamiltonian (for a review, see Ref.~\cite{souvik2021}). This winding number correctly predicts the number of the Floquet Majorana modes (zero modes and $\pi$-modes) for a Kitaev chain with a periodically driven chemical potential.
However, the dynamical winding number fails to detect the anomalous edge modes 
which arise when the hopping parameter is driven strongly~\cite{saha17}.

Recently, the notion of a disconnected entanglement entropy (DEE)~\cite{fromholz20, 
micallo20,msaikat22} has been introduced which plays a role similar to a topological invariant 
in an equilibrium Kitaev chain with an open boundary condition.
It is worth noting that unlike the winding number, the DEE is not a bulk topological 
invariant. Rather, it extracts the entanglement between the Majorana modes localized 
at the edges. Although the DEE can take any real value by its construction, it turns 
out to be integer-quantized for a short-ranged Kitaev chain in the topological 
phase, where the integer is the number of edge modes localized at each edge of the chain~\cite{fromholz20}.

In this paper, we explore the efficacy of the DEE in detecting the dynamically
generated edge modes for a periodically driven Kitaev chain. We study the variation of the DEE with the drive frequency for a periodic modulation of the chemical potential with $\delta$-function pulses. Our study
establishes that the DEE correctly predicts the number of Floquet Majorana edge modes. We also investigate the applicability of the DEE as a marker of anomalous edge modes appearing due to a periodic modulation of the nearest-neighbor hopping amplitude.

It is also noteworthy that the verification of topological properties of periodically driven systems are experimentally more viable compared to undriven systems.
Recently, a periodically driven Ising model in the presence of an additional site-dependent random longitudinal field has been implemented experimentally using an array of superconducting qubits~\cite{mi22}. It is important to note that in the integrable limit (when the longitudinal field is absent), a transverse field Ising model can be mapped to a Kitaev chain through Jordan-Wigner transformation~\cite{lieb61}. Although the Jordan-Wigner transformation is non-local, there is no non-local term consisting of long string operators in the Jordan-Wigner transformed Ising Hamiltonian with the open boundary conditions (unlike with the periodic boundary conditions) and thus it works fine with the open boundary conditions~\cite{oshikawa19}.
In the present work, we not only probe how these Floquet Majorana modes are manifested in the DEE, but also we illustrate the robustness of these modes against weak spatial disorder and temporal noise.

The rest of the paper is organized as follows: the conventional definition of the DEE is introduced in Sec.~\ref{sec_dee}. In Sec.~\ref{sec_srk}, we briefly recapitulate a short-ranged Kitaev chain of spinless fermions and its topological properties. The behavior of the DEE for a Kitaev chain with periodically modulated chemical potential is explored in Sec.~\ref{sec_drive}. We analyze the effects of spatial disorder and temporal noise in the periodic driving on the DEE in Sec.~\ref{sec_disorder}. In Sec.~\ref{sec_anomalous}, we study the situation where the nearest-neighbor hopping amplitude is periodically modulated and address the question of whether the DEE can detect the anomalous edge modes. In Sec.~\ref{sec_hx}, probing the DEE, we analyse the robustness of the Floquet Majorana modes against an integrability breaking interaction in a periodically driven Ising model, as studied experimentally in Ref.~\cite{mi22}. Concluding remarks are presented in Section~\ref{sec_conclusion}. The definition of the winding number of a short-range (static) Kitaev chain is given in Appendix~\ref{sec_topology}. The methods used for computing the DEE from Floquet Hamiltonian are explained in Appendix~\ref{app_method}.
We then briefly compare the DEE with the dynamical winding number derived from the Floquet Hamiltonian for a periodic variation of the chemical potential in Appendix~\ref{sec_winding}. In Appendix~\ref{app_gap}, we provide an analytical derivation of the frequencies at which Floquet quasienergy gap becomes zero or $\pm \pi/T$, choosing the examples of a periodic driving of the hopping amplitude and an experimentally relevant driving protocol (discussed in Sec.~\ref{sec_hx}) for an Ising chain.

\section{Disconnected entanglement entropy}
\label{sec_dee}

\begin{figure}
\centering
\includegraphics[scale=0.45]{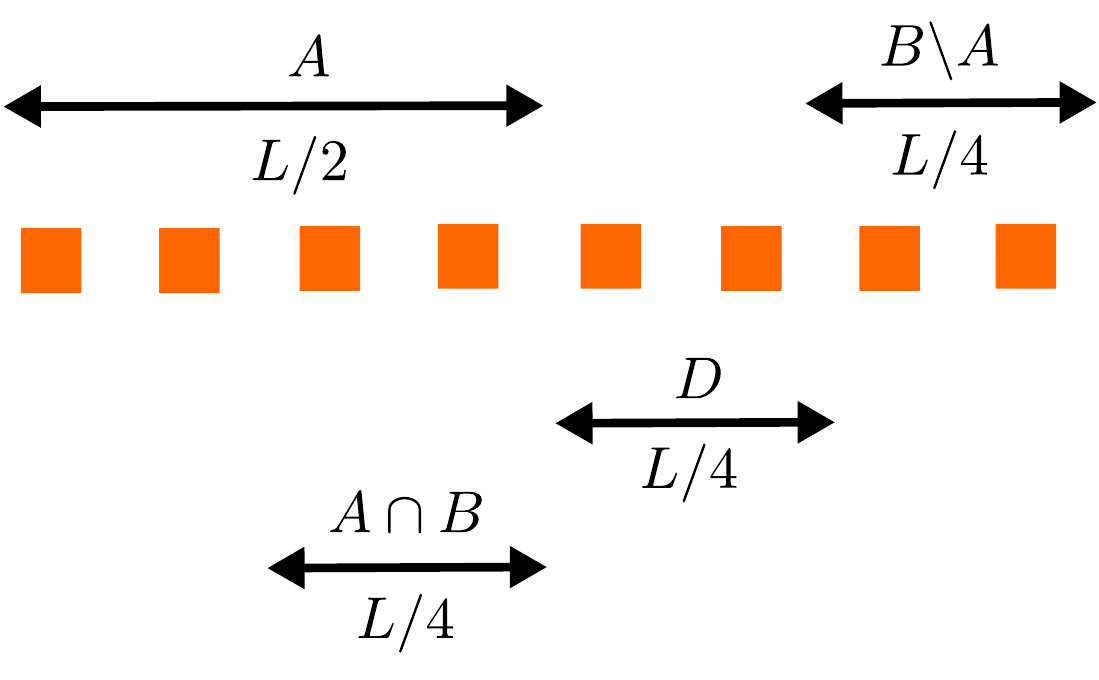}
\centering
\caption{Partitions of a chain with the disconnected partition $D =
\overline{A \cup B}$ and $2 L_{A} = 2 L_{B} = 4 L_{D} = L$. The subsystem $B$ 
consists of two partitions $A \cap B$ and $B \backslash A$, separated by the 
disconnected partition $D$. (In this figure, each square represents a fermionic site.)} \label{fig_partitions} \end{figure}

In this section, we briefly introduce the notion of disconnected entanglement 
entropy (DEE)~\cite{fromholz20, micallo20}. To this end, we first consider a 
composite system $\mathcal{S}$ that is in a pure state and is described by a density
matrix $\rho$. The reduced density matrix of a subsystem $A$ is obtained by tracing 
over the degrees of freedom of the rest of the system $\overline{A}$ \cite{peschel03}:
\begin{equation}\label{eqn_rho_A}
\rho_{A} = \Tr_{\overline{A}} (\rho). \end{equation}
The entanglement entropy~\cite{vidal03, calabrese04, latorre09, calabrese09} of 
subsystem $A$ is then defined in terms of the eigenvalues
$\lambda_i$ of the reduced density matrix $\rho_{A}$ as
\begin{equation}\label{eqn_SA}
S_{A} = - \Tr_{A} (\rho_{A} \ln (\rho_{A}) ) = - \sum _{i} \lambda_{i} \ln (\lambda_{i}).
\end{equation}
As the system $\mathcal{S}$ is in a pure state, it can be shown that
$S_{A} = S_{\overline{A}} $.

We now consider a configuration of the partitions $A$, $B$, $A \cap B$ and $A \cup B$ 
of the system $\mathcal{S}$, such that the subsystem $B$ consists of two parts $A \cap B$ 
and $B \backslash A$, separated by the disconnected partition $D=\overline{A \cup B}$, as 
shown in Fig.~\ref{fig_partitions}. The DEE~\cite{fromholz20, micallo20, msaikat22} 
is then defined as
\begin{equation}\label{eqn_SD}
S_{D} = S_{A} + S_{B} - S_{A \cup B} - S_{A \cap B}. \end{equation}

\section{Short-range Kitaev chain}
\label{sec_srk}

We recall the Hamiltonian of a Kitaev chain~\cite{Kitaev01, laumann09, DeGottardi11, DeGottardi13, dsen13} with short-ranged couplings given by
\begin{align}\label{eqn_H}
H = - \gamma \sum_{n=1}^{L-1} (c_{n}^{\dagger} c_{n+1} + c_{n+1}^{\dagger} c_{n} ) - \mu \sum_{n=1}^{L} ( 2 c_{n}^{\dagger} c_{n} -1 ) \nonumber  \\ + \Delta \sum_{n=1}^{L-1} (c_{n} c_{n+1} + c_{n+1}^{\dagger} c_{n}^{\dagger}),
\end{align}
where $c_{n}$ ($c_{n}^{\dagger}$) is the annihilation (creation) operator for a spinless 
fermion on the $n$-th site, $\gamma$ is the nearest-neighbor hopping parameter, $\Delta$ 
is the strength of the $p$-wave superconducting pairing, and $\mu$ is the on-site chemical
potential. The parameters $\gamma$, $\Delta$ and $\mu$ will be taken to be real unless
otherwise mentioned. Consequently, the Hamiltonian respects time-reversal symmetry, particle-hole symmetry and sub-lattice$/$chiral symmetry and the system belongs to the BDI symmetry class~\cite{dsen13,trs_sen13}. Definition of the 
winding number and topological properties of a short-range Kitaev chain are briefly 
discussed in Appendix~\ref{sec_topology}.

It is noteworthy that in a chain with short-ranged couplings and open boundary 
conditions, the DEE, calculated in the ground state of static Hamiltonian, is an 
integer multiple of $\ln(2)$, i.e., $S_{D}=p \ln(2)$, where $p$ is the number 
of modes localized at each edge of a chain with the open boundary conditions (see Ref.~\cite{fromholz20}).
Therefore, the values of the DEE in the topologically non-trivial ($-1<\mu<1$) and trivial ($|\mu| >1$) phases of a Kitaev chain with $\gamma=1$ and $\Delta \neq 0$ are $\ln(2)$ and
zero, respectively, and there is a discontinuous jump in the value of the DEE at the
QCP separating the two phases. Thus, the DEE plays a role equivalent to the 
winding number for an open chain.

\section{DEE for a Kitaev chain with periodically modulated chemical potential}
\label{sec_drive}
\begin{figure}
\centering
\includegraphics[width=0.45\textwidth]{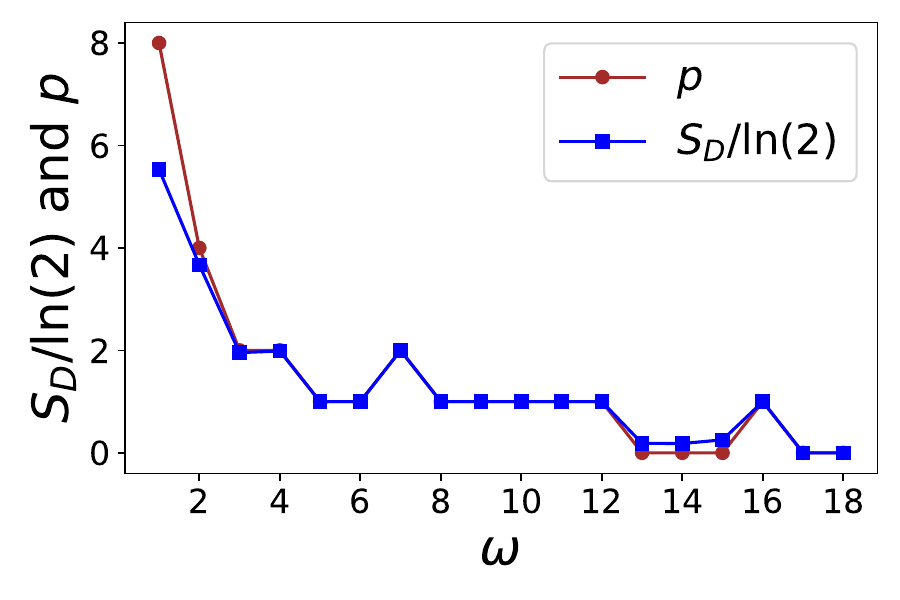}
\centering	
\caption{DEE ($S_{D}$) in units of $\ln (2)$ and the number of Majorana 
edge modes ($p$) localized at each edge as functions of driving frequency 
$\omega$ (in the units of the hopping parameter $\gamma$, which is set equal to unity) for a Kitaev chain in which the chemical potential $\mu$ is periodically modulated by $\delta$-pulses
(Eq.~\eqref{eq_delta}) with $\mu_{0}=2.5$ and $\mu_{1}=0.2$. We have taken $\Delta=1$, $L=200$ and
$L_{D}=50$. Here, both the parameters $\mu_{0}$ and $\Delta$ are in the units of $\gamma$, while the parameter $\mu_{1}$ is dimensionless. The DEE assumes integer-quantized values except at low frequencies.
The number of dynamically generated edge Majorana modes increases as $\omega$ decreases.} \label{fig_delta} \end{figure}


In this section, we discuss the behavior of the DEE calculated in the ground state of the effective Floquet Hamiltonian for a Kitaev chain with a periodically modulated chemical potential.

We consider a Kitaev chain with an open boundary condition in which the 
chemical potential is periodically modulated~\cite{dsen13}, such that $\mu (t) = \mu 
(t+T)$, with $T=2\pi/\omega$, where $\omega$ is the driving frequency, so that
$H(t)$ in Eq.~\eqref{eqn_H} satisfies $H(t)=H(t+T)$.
For a time-periodic Hamiltonian $H(t)$, the stroboscopic time-evolution operator 
(i.e., the Floquet operator) is defined as
\begin{multline}\label{eqn_U}
U_{F}=\mathbb{T} \exp \left(-i \int_{0}^{T} H (t) dt \right) = \exp(-iH_{F}T),
\end{multline}
where $H_{F}$ is the Floquet Hamiltonian and $\mathbb{T}$ denotes time-ordering. 
(We will set $\hbar =1 $ in this paper).

We recall that due to the unitary nature of the Floquet operator $U_{F}$, its 
eigenvalues
are phases. Further, these appear in complex conjugate pairs, $e^{i\theta}$ and 
$e^{-i\theta}$ (see Appendix~\ref{app_method} for details) because of particle-hole symmetry (as the system belongs to Floquet BDI symmetry class~\cite{dsen13}). If present, Majorana edge modes correspond to the eigenvalues $+1$ and $-1$ of the Floquet operator ($U_{F}$). In other words, their corresponding Floquet quasienergies are $\epsilon_{F}=0$ and 
$\epsilon_{F}=\pm \pi/T$, respectively. Thus, the total number of Majorana 
edge modes (twice the number of Majorana modes at each edge) is given by the total number of eigenvalues $+1$ and $-1$ of $U_{F}$.\\

\begin{figure}
\centering
\includegraphics[width=0.45\textwidth]{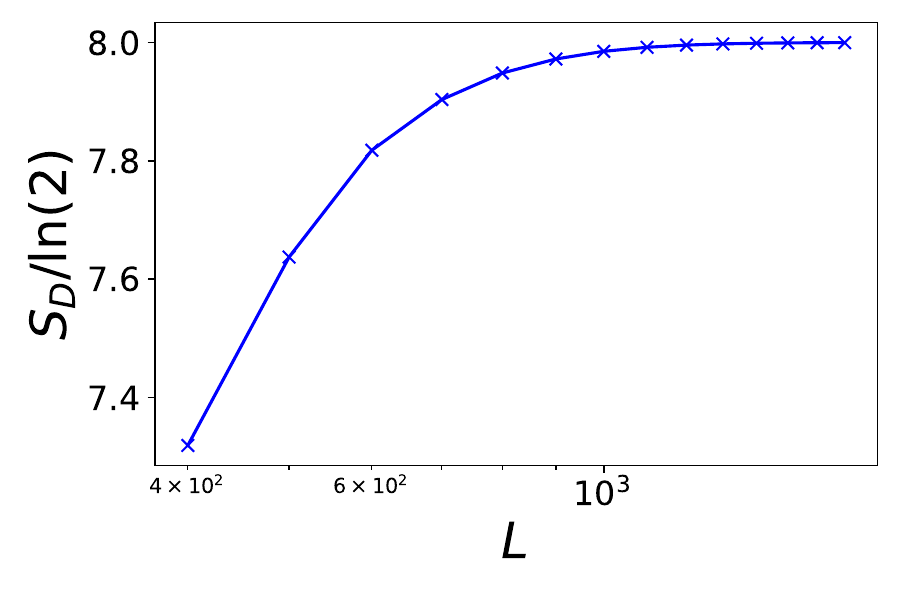}
\centering	
\caption{DEE ($S_{D}$) in units of $\ln (2)$ as a function of the length $L$ (in semi-log scale) of a
Kitaev chain (with $L_{D}=L/4$) in which the chemical potential $\mu$ is periodically modulated by $\delta$-pulses (Eq.~\eqref{eq_delta}) with
$\mu_{0}=2.5$, $\mu_{1}=0.2$, $\Delta=1$ and $\omega=1$. Here, $\mu_{0}$, $\omega$ and $\Delta$ are in the units of $\gamma$, while the parameter $\mu_{1}$ is dimensionless. We
observe a saturation to an integer-quantized value as $L$ increases, 
establishing that 
the discrepancy observed for low $\omega$ in Fig.~\ref{fig_delta} is a finite-size 
effect.} \label{fig_L} \end{figure}

We now consider the protocol in which the chemical potential $\mu$ is periodically 
modulated by the application of $\delta$-pulses such that~\cite{dsen13, souvik2021}
\begin{equation}\label{eq_delta}
\mu (t) = \mu_{0} + \mu_{1} \sum_{m=-\infty}^{\infty} \delta ( t -m T ),
\end{equation}
where $T=2 \pi/ \omega$. Both $\omega$ and $\mu_{0}$ are in the units of the hopping parameter $\gamma$, where we have set $\gamma=1$. On the other hand, the parameter $\mu_{1}$ is dimensionless. \\

The chemical potential $\mu(t)$ in Eq.~\eqref{eq_delta} for a Kitaev chain can be mapped to a transverse field in an Ising chain through the Jordan-Wigner transformations~\cite{lieb61}. If the transverse field of an Ising chain is periodically kicked (with $\delta$-function kicks), then the driving protocol of the quantum Ising chain can be mapped to the driving protocol as in Eq.~\eqref{eq_delta}. It is noteworthy that a quantum Ising model can be simulated with quantum circuits~\cite{li14} and $\delta$-function 
 kicks can be experimentally realized with laser pulses~\cite{raizen95}.

We now calculate the DEE in the ground state of the Floquet Hamiltonian $H_{F}$ for this driving protocol in a Kitaev chain with open boundary conditions. The methods used to calculate the DEE from the correlation matrix are explained in Appendix~\ref{app_method}. In this driving protocol, the DEE, calculated in the ground state of the Floquet Hamiltonian, is equal to integer multiples of $\ln(2)$, as can be seen from Fig.~\ref{fig_delta}. Thus, $S_{D}=p\ln(2)$, where $p$ is an integer. It is also interesting to note that the DEE generally increases as the drive frequency $\omega$ decreases. Further, we have verified that the value of $S_{D}/\ln(2)$ is equal to the number of Majorana edge modes (both zero and $\pi$-modes) localized at each edge of the chain (see Fig.~\ref{fig_delta}) generated by the same periodic driving of the Kitaev chain. 

However, at low drive frequencies (i.e, when $\omega$ is of the order of the hopping parameter $\gamma$), the DEE is not integer-quantized and the value of $S_{D}/\ln(2)$ differs significantly from the number of Majorana end modes. This is however a finite-size effect, as shown in Fig.~\ref{fig_L}, which demonstrates that the DEE does saturate to an integer-quantized value for large system size $L$. The reason for the finite-size effect is that at low frequencies, the decay lengths of the Majorana edge modes increase and therefore the edge modes mix more and more with the bulk states. In other words, the average normalized participation ratio (NPR) for the edge-modes is of the order of unity at low drive-frequencies~\cite{tilen19}, while the NPR is of the order of $1/L$ when the frequency is not sufficiently low~\cite{tilen19}. (For a normalized wave function $\psi_j (n)$, where $j$ labels the eigenstate and $n$ denotes the site index, the participation ratio (PR) is defined as $R_{j}=1/(\sum_n |\psi_j (n)|^4)$ and the average normalized participation ratio (NPR)~\cite{tilen19} is defined as, NPR$= \langle R_{j} \rangle /L$, where the average is taken over all the eigenstates. The average NPR is of the order of unity and $1/L$ for extended states and localized states, respectively~\cite{tilen19}.) When the decay length is comparable to the size $L/4$ of the disconnected region $D$, the contribution of such end modes to the DEE deviates from integer multiples of $\ln (2)$ (i.e., the DEE fails to detect these edge-modes correctly). This leads to the reduction in the contribution of the edge-modes to the DEE. On the other hand, at low frequencies, the bulk states become long-range entangled (as the effective range of couplings appearing in the Floquet Hamiltonian increases with decreasing drive frequency) and the long-range entangled bulk states contribute to the DEE (see Ref.~\cite{msaikat22}). This effect leads to an increase in the DEE. For small $\omega$, therefore, the deviation of the DEE from integer-quantized value occurs due to the competition between these two contradictory effects. As the length $L_{D}=L/4$ is increased, the bulk contribution to the DEE decreases~\cite{msaikat22}. Also, with the increase of $L$, the dispersion of the edge modes to the bulk states decreases (as the decay lengths of the Floquet Majorana modes are much smaller than the system-size $L$, when $L$ is sufficiently large), even at low drive frequencies. Thus, for sufficiently large $L$ (and therefore sufficiently large $L_{D}=L/4$), the bulk contribution becomes negligibly small and the DEE only contains the integer-quantized contribution of the edge modes.\\

The DEE is found to be equivalent to the dynamical (Floquet) winding number 
in counting the number of emergent Floquet Majorana modes; this is discussed in Appendix~\ref{sec_winding}.

\section{DEE for periodic modulation of chemical potential in presence of spatial disorder and temporal noise}\label{sec_disorder}

The Majorana edge-modes are robust against weak perturbations and the robustness of the Majorana modes against weak perturbations has been explored in several studies (see Ref.~\cite{hu15,balabanov17,rieder18,jorg19}). In this section, we discuss the behavior of the DEE for a periodically driven Kitaev chain in the presence of spatial disorder as well as temporal noise, and we establish that the Floquet Majorana modes are robust against weak spatial and temporal disorder.

\subsection{DEE for periodic driving in presence of spatial disorder}\label{sec_spatial}
\begin{figure}
\centering
\includegraphics[width=0.45\textwidth]{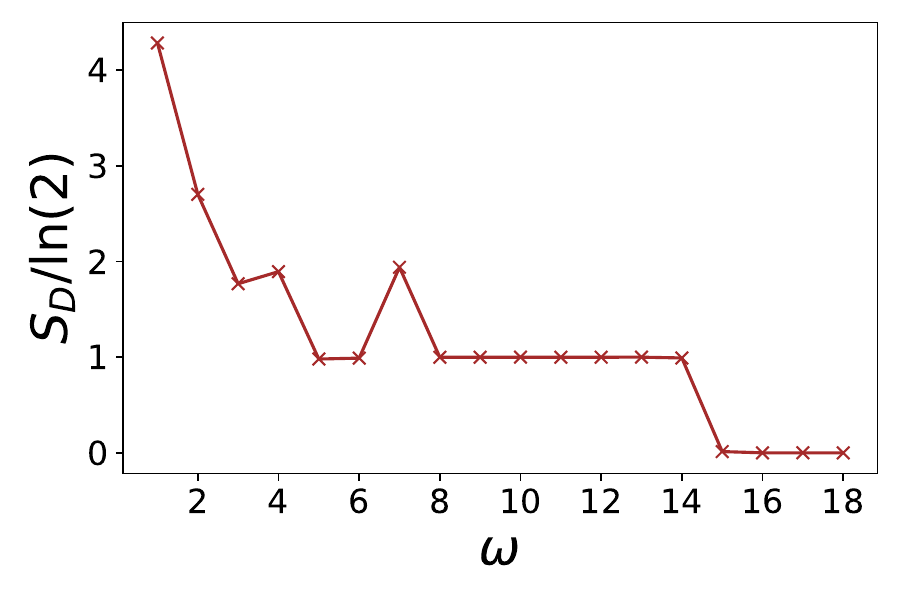}
\centering
\caption{DEE ($S_{D}$) in units of $\ln(2)$ as a function of $\omega$ (in the units of the hopping parameter $\gamma$, which is set equal to unity) for a driving protocol in the presence of spatial disorder as given in Eq.~\eqref{eq_mu_spatial}. We have chosen $\mu_{0}=2.5$, $r=0.2$, $\Delta=1$, $L=200$ and $L_{D}=50$. Here, $\mu_{0}$ and $\Delta$ are in the units of $\gamma$, while the parameter $r$ is dimensionless. The dimensionless parameter $\beta_{n}$ assumes a random value in the range $[0,1]$. Here, 
$S_{D}$ has been calculated after averaging over a large number of configurations of random numbers.} \label{fig_spatial} \end{figure}

We consider a driving protocol where the on-site potentials $\mu_{n}(t)$ for different sites are driven with $\delta$-pulses of random amplitudes, such that the translation symmetry of the chain is explicitly broken. Therefore, we consider the following form of the chemical potential,
\begin{equation}\label{eq_mu_spatial}
\mu_{n} (t)= \mu_{0} + r \beta_{n} \sum_{m=-\infty}^{\infty} \delta(t-mT),
\end{equation}
where $n$ denotes the site number in the chain, $\mu_{0}$ and $r$ are independent of $n$, and $\beta_{n}$ can assume a random value in the range $[0,1]$. Here, both the parameters $r$ and $\beta_{n}$ are dimensionless.

After averaging over a large number of configurations of random numbers, we find 
that the value of the DEE, calculated in the ground state of the effective Floquet 
Hamiltonian, remains perfectly integer-quantized even in the presence of weak spatial disorder for high frequency (see 
Fig.~\ref{fig_spatial}). This demonstrates the robustness of Floquet Majorana modes 
(both zero and $\pi$-modes) against weak disorder. However, the deviation of the DEE from integer-quantized value at sufficiently low-frequency is due to the finiteness of the system, as we have elaborated in Fig.~\ref{fig_L} for the driving protocol without spatial disorder.

\subsection{DEE for periodic driving in presence of temporal noise}\label{sec_noise}
\begin{figure}
\centering
\includegraphics[width=0.45\textwidth]{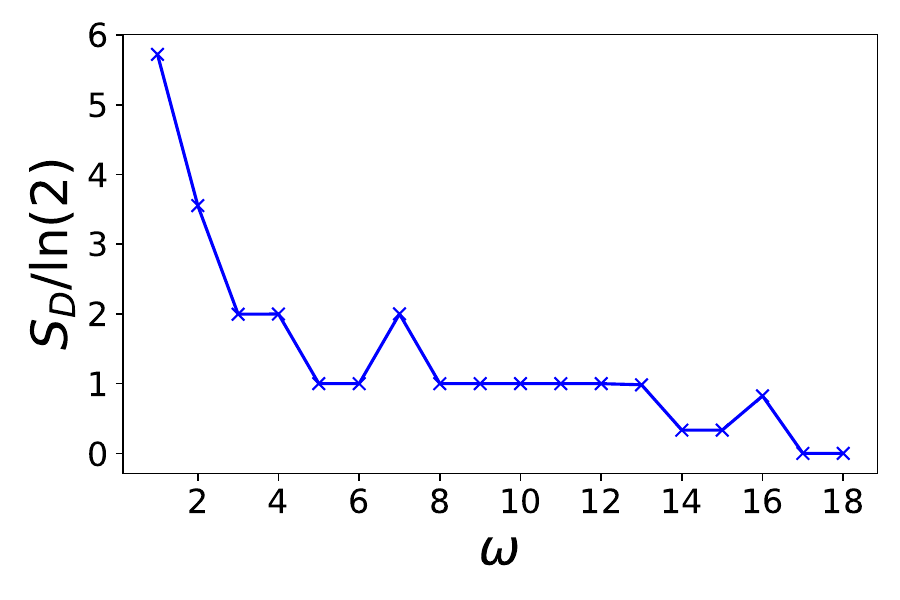}
\centering
\caption{DEE ($S_{D}$) in units of $\ln(2)$ as a function of $\omega$ (in the units of the hopping parameter $\gamma$, which is set equal to unity) for a driving protocol in the presence of temporal noise as 
given in Eq.~\eqref{eq_mu_noise}. We have chosen $\mu_{0}=2.5$, $\mu_{1}=0.2$, $r=0.2$, $\Delta=1$, $L=200$ and $L_{D}=50$. Here, $\mu_{0}$ and $\Delta$ are in the units of $\gamma$, while the parameters $\mu_{1}$ and $r$ are dimensionless. $f(t)$ assumes a random value in the range $[-1,1]$ in the units of $\gamma$ at any time $t$ lying in the range 
$[0,T]$, where $T=2\pi/\omega$.} \label{fig_noise} \end{figure}

We now consider a driving protocol where the chemical potential $\mu(t)$ is spatially uniform, but is driven with $\delta$-pulses in the presence of a random noise $f(t)$ of sufficiently small amplitude $r$. We consider the following form of the chemical potential,
\begin{equation}\label{eq_mu_noise}
\mu (t)= \mu_{0} + \mu_{1} \sum_{n=-\infty}^{\infty} \delta(t-nT) + r f(t),
\end{equation}
where $T=2\pi/\omega$ and the function $f(t)$ assumes a random value between $[-1,1]$ for $0<t<T$. \tb{Here, $r$ is a dimensionless parameter and the value of $f(t)$ is in the units of the hopping parameter $\gamma$, where we have set $\gamma=1$.} However, we still consider the function $f(t)$ to be time-periodic with period $T$, i.e., $f(t)=f(t+T)$. The time-periodicity of the function $f(t)$ renders the Hamiltonian $H(t)$ periodic in time, i.e., $H(t)=H(t+T)$. We compute the DEE in the ground state of the Floquet Hamiltonian $H_{F}$ for the protocol as in Eq.~\eqref{eq_mu_noise}.

For this driving protocol with a temporal noise, we find that the value of the DEE remains 
invariant as shown in Fig.~\ref{fig_noise}. This indicates 
the robustness of the dynamically generated Floquet Majorana modes against a
weak temporal noise in the driving.

\section{Detection of the anomalous edge modes with periodically modulated hopping parameter through DEE}
\label{sec_anomalous}

\begin{figure*}
\centering
\subfigure[]{
\includegraphics[width=0.45\textwidth]{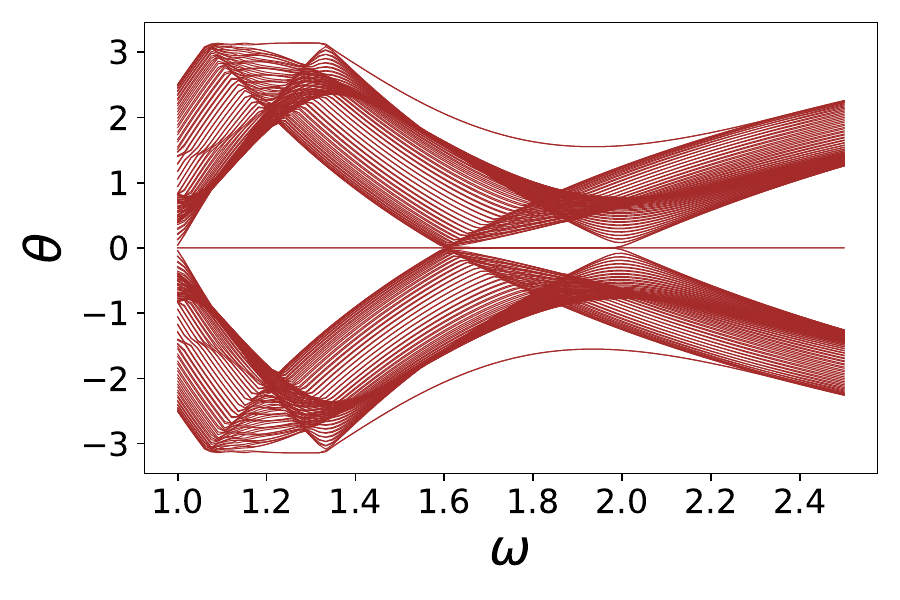}
\label{fig_sp_a1_sym}}	
\centering
\hspace{5mm}
\subfigure[]{
\includegraphics[width=0.45\textwidth]{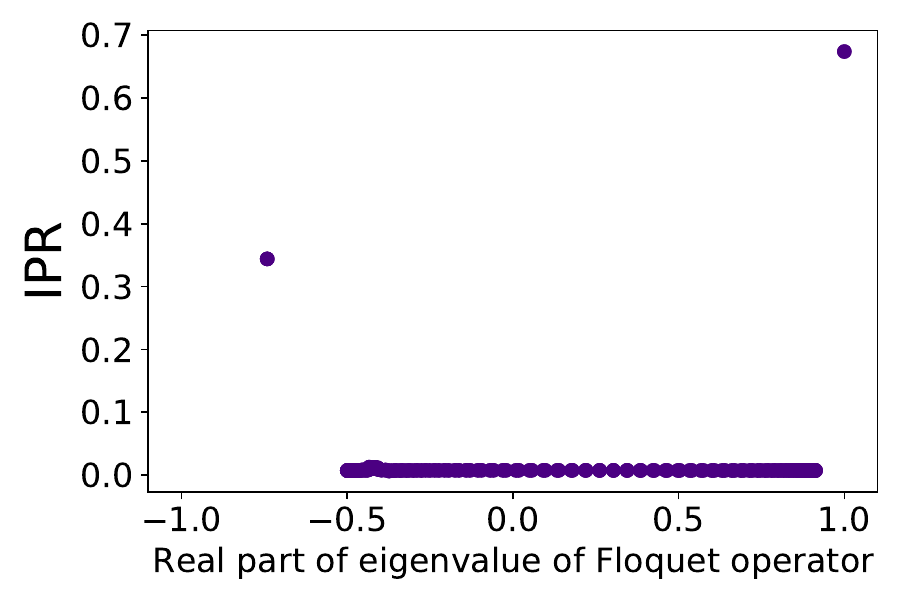}
\label{fig_ipr_a1_sym}}
\subfigure[]{
\includegraphics[width=0.45\textwidth]{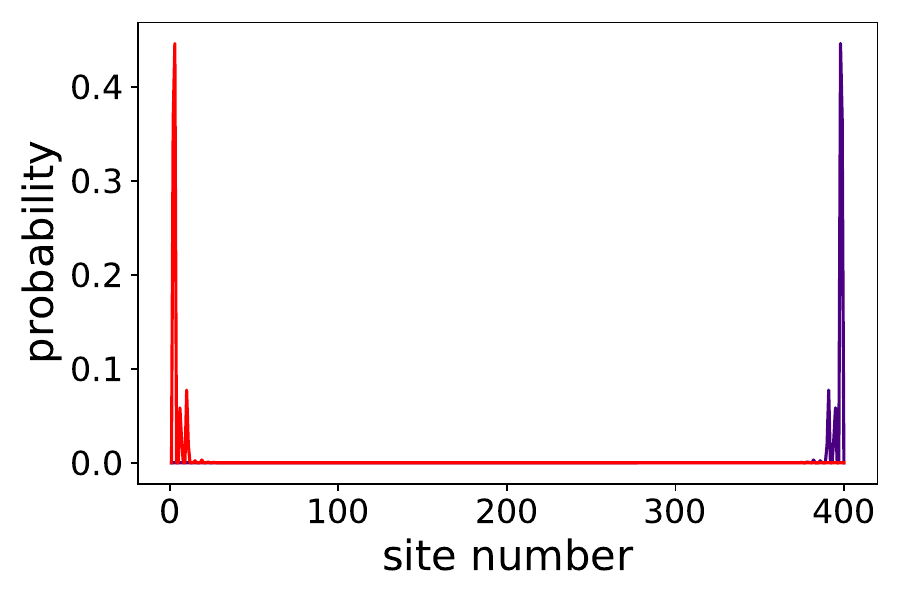}
\label{fig_prob_a1_sym}}	
\subfigure[]{
\includegraphics[width=0.45\textwidth]{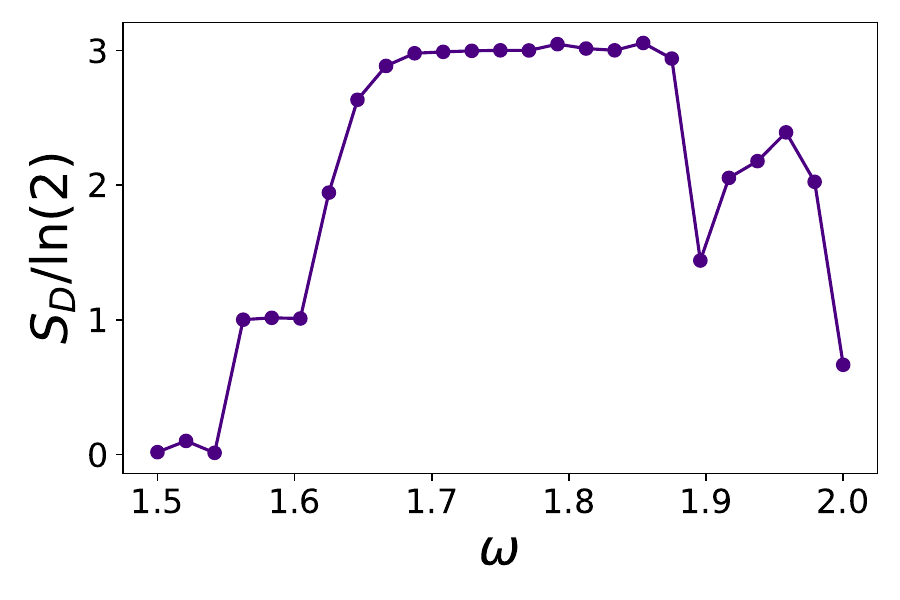}
\label{fig_sd_sym}}
\subfigure[]{
\includegraphics[width=0.45\textwidth]{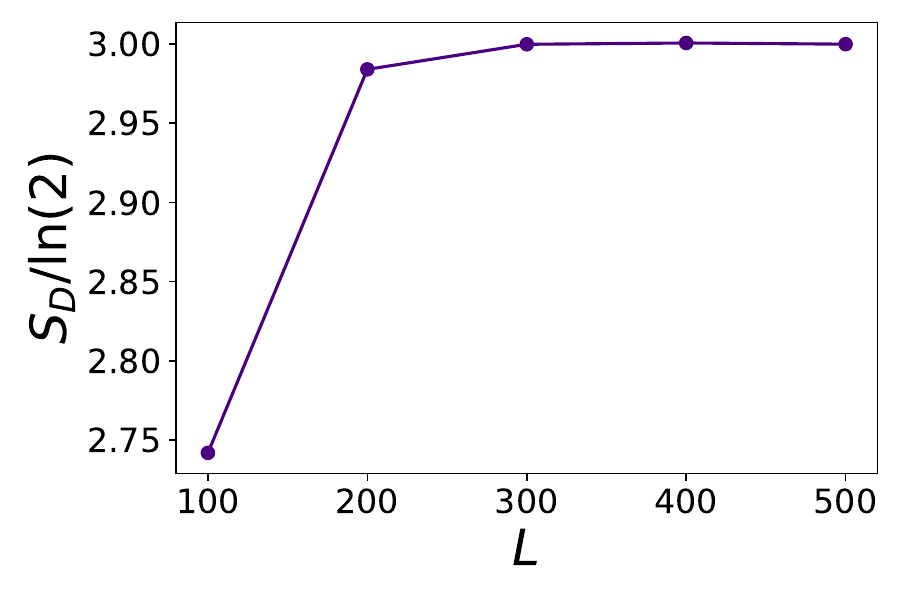}
\label{fig_sym_L}}
\caption{(a) Floquet quasienergies $\theta$ as a function of the drive frequency 
$\omega$ \tb{(in the units of $\gamma_{0}$, which is set equal to unity)}). The two isolated lines seen near the top and bottom in the 
frequency range of about $[1.4,2.2]$ correspond to anomalous end modes. The two isolated lines seen near the top and bottom in the 
frequency range of about $[1.1,1.4]$ correspond to Majorana $\pi$ modes.
(b) Inverse participation ratio (IPR) as a function of the real part of 
the eigenvalue of the Floquet operator $U_{F}$ at $\omega=1.6$ and for $L=200$. The modes with the real part of eigenvalue $+1$ of the Floquet operator are Majorana zero modes and the modes with the minimum real part of eigenvalue (if it is not equal to $-1$) of the Floquet operator are anomalous modes. There are four anomalous modes (two anomalous modes at each edge) and two zero-energy modes (one zero-energy mode at each edge). (c) Probability as a function of the site number for the two anomalous modes with the eigenvalues $-0.7412 \pm 0.6713 i$ of the Floquet operator $U_{F}$ at $\omega=1.6$. (d) DEE ($S_{D}$) in units of $\ln (2)$
as a function of $\omega$ for $L=400$ and $L_{D}=100$. \tb{(e) DEE ($S_{D}$) in units of $\ln (2)$ as a function of the system-size $L$ at $\omega=1.7$. The DEE saturates to the value $3 \ln(2)$ at $\omega=1.7$ for sufficiently large $L$.} For all the plots, the amplitude of the hopping parameter is periodically modulated (see Eq.~\eqref{eq_sym}) with \tb{the dimensionless parameter} $a=1.0$. We have taken $\mu=0$ and $\Delta=0.8$; \tb{both $\mu$ and $\Delta$ are in the units of $\gamma_{0}$, where we have set $\gamma_{0}=1$.}} \label{fig_sym} \end{figure*}

We now consider the periodic driving of the hopping parameter $\gamma$, which may 
be complex in general, so that the Hamiltonian in \eqref{eqn_H} is modified to
\begin{align}\label{eqn_H_trs}
H (t) = - \sum_{n=1}^{L-1} ( \gamma (t) c_{n}^{\dagger} c_{n+1} + \gamma^{\star} (t) c_{n+1}^{\dagger} c_{n} ) \nonumber \\ - \mu \sum_{n=1}^{L} ( 2 c_{n}^{\dagger} c_{n} -1 ) + \sum_{n=1}^{L-1} \Delta ( c_{n} c_{n+1} + c_{n+1}^{\dagger} c_{n}^{\dagger}).
\end{align}
 It has been established that for this type of driving protocol, the modes localized 
 at the edges with the eigenvalues of the Floquet operator away from $\pm 1$ (referred to as anomalous
 modes) can dynamically emerge~\cite{saha17}. We address here the question of whether 
 the DEE can detect these anomalous modes.

\subsection{DEE for periodic driving of the hopping parameter}
\label{subsec_sym}

We consider the following form of the hopping parameter,
\begin{equation}\label{eq_sym}
\gamma(t)= \gamma_{0} \left( 1+ a \cos (\omega t) \right), \end{equation}
where $\omega=2\pi/T$. \tb{Here, $\omega$ is in the units of $\gamma_{0}$, where we have set $\gamma_{0}=1$.} \tb{The dimensionless parameter} $a$ determines the strength of the modulation of the hopping amplitude.
Since the hopping parameter is chosen to be real, the time-reversal symmetry of the 
Hamiltonian $H(t)$ is preserved, and the periodically driven chain belongs to the BDI 
(Floquet) symmetry class~\cite{dsen13,trs_sen13,saha17}.

To see if there are emergent anomalous modes, we plot the Floquet quasienergies $\theta$ as a
function of $\omega$ for $a=1$ in Fig.~\ref{fig_sp_a1_sym}. Evidently, for $1.5<\omega<2.0$, the 
extreme values (near the top and bottom) of the Floquet quasienergies are separated by finite gaps
from the other quasienergies. The modes with these isolated values of Floquet quasienergies are 
known as anomalous modes~\cite{saha17} since the corresponding eigenvalues of the Floquet operator 
differ from $\pm 1$.

In Fig.~\ref{fig_ipr_a1_sym}, the variation of the inverse participation ratio (IPR) is plotted against the real part of the corresponding eigenvalues of the Floquet operator for $\omega=1.6$. (For a normalized wave function $\psi_j (n)$, where $j$ labels the wave function and $n$ denotes the site index, the IPR is defined as $\sum_n |\psi_j (n)|^4$. It is known that as the system size $L$ is taken to infinity, the IPRs of modes which are extended in the bulk go to zero while the IPRs of modes localized at the ends remain finite. Hence a plot of the IPR versus $j$ provides an easy way to identify the edge modes). From this figure, it can clearly be seen that the anomalous modes [which have the minimum real part (not equal to $-1$) of the eigenvalues of the Floquet operator] have relatively large IPRs, compared to the other modes having non-zero Floquet quasienergies. We note here that the IPRs of the anomalous modes are almost comparable to that of the zero-energy Majorana modes (with the eigenvalue of the Floquet operator being equal to $+1$). Fig.~\ref{fig_prob_a1_sym} further confirms that these anomalous modes, despite having non-zero Floquet quasienergies, are localized at the edges of the chain~\cite{saha17}.

In Fig.~\ref{fig_sd_sym}, the DEE ($S_{D}$) is plotted as a function of the 
drive frequency $\omega$ to find the contribution of both the anomalous modes and the zero-energy modes to the 
DEE. As is evident from Fig.~\ref{fig_sd_sym}, in the approximate range of frequencies between $[1.6,1.8]$, the non-zero and integer-quantized 
contribution of the anomalous modes and the zero-energy modes to the DEE enables us to detect these edge
modes. In this range of frequencies, there are two anomalous modes and one Floquet zero-energy Majorana mode at each edge of the chain (one zero-energy Majorana mode at each edge is generated at $\omega=2$ for $\gamma_{0}=1$; as the Floquet quasienergy gap closes at $\omega=2$ for $\gamma_{0}=1$, which can be seen from the analytical calculation presented in Appendix~\ref{app_hgap}) and the DEE is nearly equal to $3\ln(2)$. \tb{Further, it can be observed from Fig.~\ref{fig_sym_L} that the DEE converges to the value $3\ln(2)$ for sufficiently large system-size $L$ for the drive frequency $\omega=1.7$.} Thus, we observe that $S_{D}/\ln(2)$ matches 
exactly with the number of edge modes (the zero-energy Majorana modes as well as the anomalous modes) localized at each edge for $1.6 \lesssim \omega \lesssim 1.8$. In this range of frequency, the DEE is able to detect the anomalous modes and zero-energy Majorana modes. Subsequently, 
we will illustrate in Sec.~\ref{sec_an_disorder} that the zero-energy Floquet Majorana modes are robust against weak spatial disorder while the anomalous modes are not.

Next, we consider a driving protocol in which the hopping parameter is complex~\cite{saha17}, with the form
\begin{equation}\label{eq_brk}
\gamma(t)= \gamma_{0} \exp \left[ i a \cos (\omega t) \right], \end{equation}
where $\omega=2\pi/T$, and the parameter $a$ determines the strength of the 
modulation of the phase of the hopping parameter. The complex hopping parameter in
Eq.~\eqref{eq_brk} explicitly breaks the time-reversal symmetry of 
the Hamiltonian $H(t)$~\cite{trs_sen13,saha17}. Therefore, a chain periodically driven with this protocol belongs to the D (Floquet) symmetry class~\cite{saha17}. For this driving protocol, similar to the driving protocol in Eq.~\eqref{eq_sym}, zero-energy Floquet Majorana modes and anomalous modes can appear. However, we find that the DEE, calculated using the methods provided in Appendix~\ref{app_method}, is not able to count the anomalous edge modes and zero-energy Majorana modes correctly for this driving protocol. \tb{In the symmetry class D, time-reversal symmetry is broken and the Majorana edge-modes are not protected from hybridization. Due to the hybridization of the edge-modes with each other (preserving particle-hole symmetry),
we find that the value of $S_{D}/\ln(2)$ is different from the total number of Majorana modes and anomalous modes localized at each end of the chain for the symmetry class D.}\\

\subsection{Effect of spatial disorder on the DEE for periodic driving of hopping parameter}\label{sec_an_disorder}

\begin{figure*}
\centering
\subfigure[]{
\includegraphics[width=0.45\textwidth]{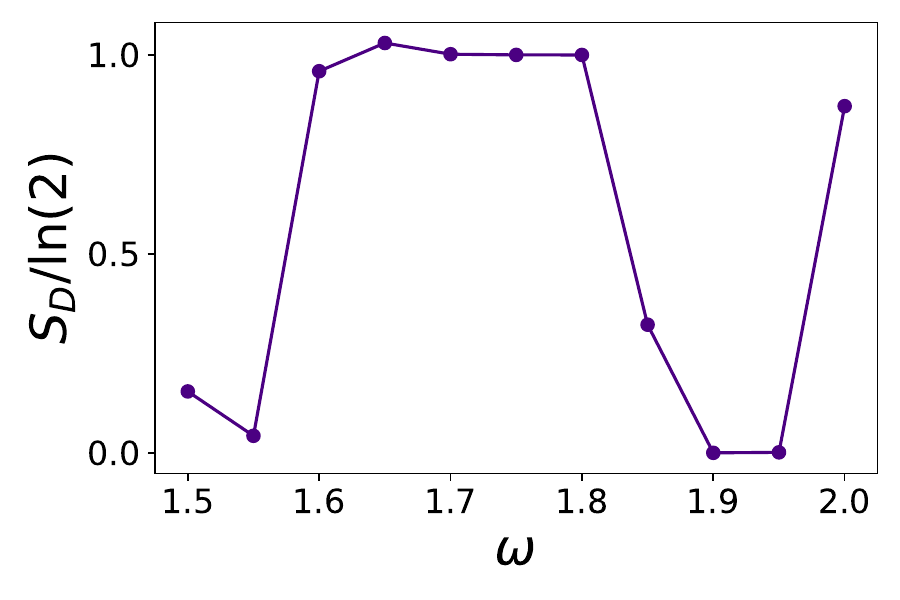}
\centering
\label{fig_sym_disorder_dee}}	
\hspace{5mm}
\centering
\subfigure[]{
\includegraphics[width=0.45\textwidth]{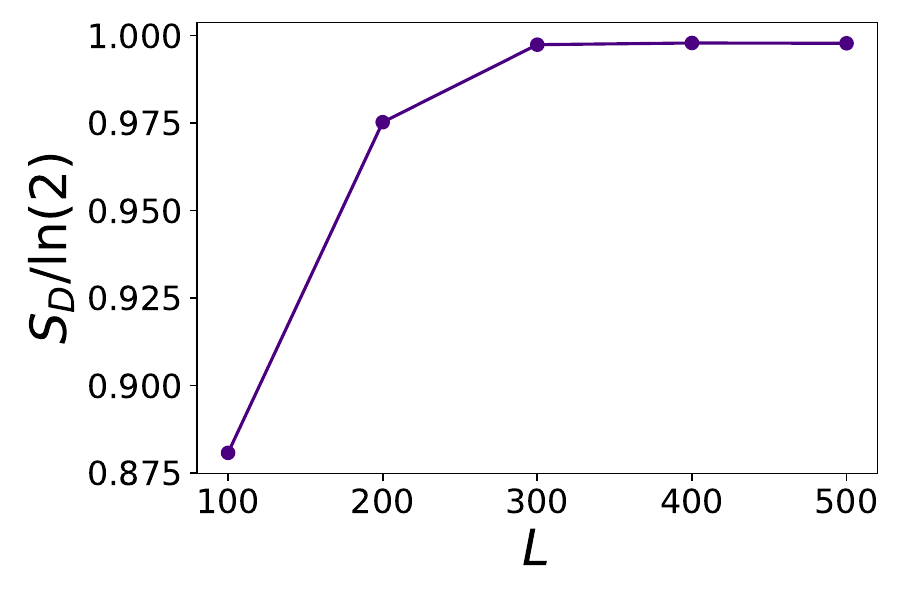}
\centering
\label{fig_sym_disorder_L}}
\caption{(a) DEE ($S_{D}$) in units of $\ln (2)$ as a function of drive frequency $\omega$ \tb{(in the units of $\gamma_{0}$, where $\gamma_{0}=1.0$)} for periodic driving of the amplitude of the hopping parameter in the presence of spatial disorder (see Eq.~\eqref{eq_sym_disorder}). We have taken $L=200$, $L_{D}=50$. \tb{(b) DEE ($S_{D}$) in units of $\ln (2)$ as a function of the system-size $L$ at $\omega=1.7$ for periodic driving of the amplitude of the hopping parameter in the presence of spatial disorder (see Eq.~\eqref{eq_sym_disorder}). The DEE converges to the value $\ln(2)$ for sufficiently large $L$.} For both the plots, we have taken $\Delta=0.8$, $\mu=0$ and $a=1.0$. \tb{Here, $\mu$ and $\Delta$ are in the units of $\gamma_{0}$, while the parameter $a$ is dimensionless.} For this protocol, \tb{the dimensionless parameter} $\nu_{n}$  assumes a random value in the range $[0,1]$ at each site $n$. The DEE has been calculated after averaging over a large number of configurations of random numbers.} \label{fig_sym_disorder}	 \end{figure*}

In this section, we investigate whether the dynamical zero-energy Majorana modes and the anomalous modes discussed in Sec.~\ref{subsec_sym} survive in the presence of weak spatial disorder. To this
end, we explore the effects of disorder on the DEE for a periodic driving of the nearest-neighbor hopping amplitude.

Let us consider the driving protocol where the amplitude of the nearest-neighbor hopping parameter $\gamma_{n}(t)$ for the $n$-th site is periodically modulated in 
the presence of spatial disorder. Therefore, this is the same driving protocol as in Eq.~\eqref{eq_sym}, but in the presence of spatial disorder. For this driving protocol, $\gamma_{n}(t)$ is given by,
\begin{equation}\label{eq_sym_disorder}
\gamma_{n}(t)=\gamma_{0} (1+ a \nu_{n} \cos(\omega t)),
\end{equation}
where $\gamma_{0}$ and the modulation strength $a$ are independent of site index $n$, and \tb{the dimensionless parameter} $\nu_{n}$ assumes random value in the range $[0,1]$ for each site $n$. From Fig.~\ref{fig_sym_disorder}, it can clearly be seen that for $1.6 \lesssim \omega \lesssim 1.8$, the DEE is almost constant and equal to $\ln(2)$ in the presence of spatial disorder. \tb{Further, it can be observed from Fig.~\ref{fig_sym_disorder_L} that the DEE saturates to the value $\ln(2)$ at $\omega=1.7$ for sufficiently large system-size $L$ for the driving protocol in the presence of spatial disorder.} On the other hand, in the absence of disorder, the DEE is equal to $3\ln(2)$ (as there are two anomalous modes and one zero-energy mode at each edge of the chain) in the same range of $\omega$ (see Fig.~\ref{fig_sd_sym}). From this observation, we infer that the zero-energy modes remain robust against spatial disorder, whereas the anomalous edge modes disappear in the presence of spatial disorder.\\

\section{DEE for experimentally studied kicked Ising model in presence of integrability breaking interactions}\label{sec_hx}
\begin{figure}
\centering
\includegraphics[width=0.45\textwidth]{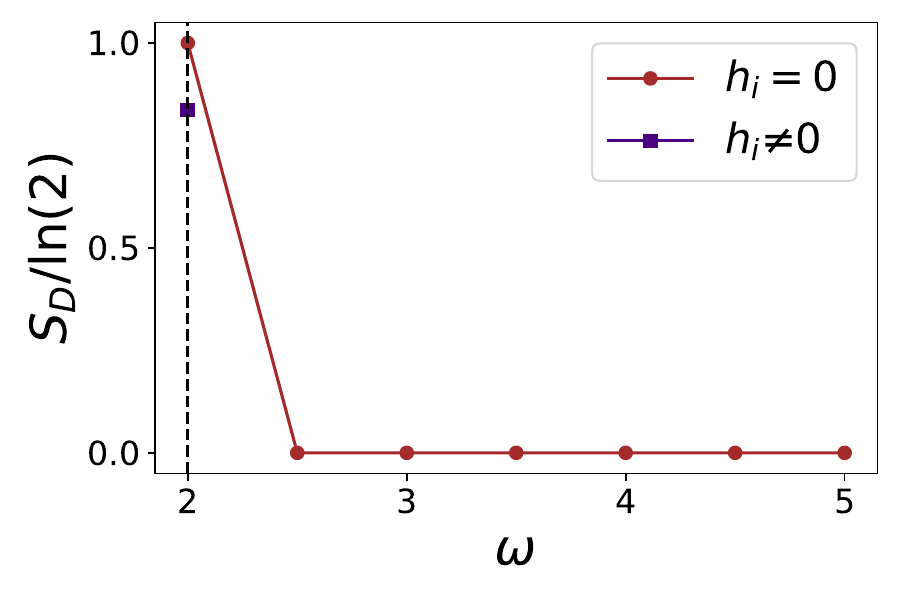}
\centering	
\caption{ {DEE ($S_{D}$) in units of $\ln (2)$ as a function of the drive frequency $\omega$ (in the units of $J$, where we have set $J=0.5$) for the driving protocol as given 
in Eq.~\eqref{eq_ilf}. The plot for $h_{j}=0$ (the integrable case) corresponds to the driving protocol as in Eq.~\eqref{eq_tfih} with the parameters $J=0.5$, $g=0.6$, $L=100$, and $L_{D}=L/4=25$. The non-integrable case with $h_{j}\neq 0$ corresponds to the presence of a site-dependent, random, longitudinal field in the driving, where the dimensionless parameter $h_{j}$ assumes a random value between $-\pi$ and $\pi$, for all $j= 1, 2, \cdots, L$.
Other relevant parameters chosen (for $h_{i} \neq 0$ and $\omega=2.0$) are $J=0.5$, $g=0.6$, $L=12$, and $L_{D}=L/4=3$. Here, the parameter $g$ is in the units of $J$. At $\omega=2.0$ (i.e., $T=\pi$), we observe that the DEE in the presence of a random longitudinal field and in the absence of the longitudinal field are nearly equal and close to $\ln(2)$. The deviation of the DEE from perfectly integer-quantized values in the presence of the longitudinal field is due to finite system size $L$, as explained in Sec.~\ref{sec_drive}.}} \label{fig_hx} \end{figure}

In this section, we discuss a recent experimental study (see Ref.~\cite{mi22}) of a kicked Ising model and we thus explore the fate of the DEE in the presence of
an integrability breaking perturbations.\\

Following Ref.~\cite{mi22}, we consider the following driving protocol of the Hamiltonian $H_{0}(t)=H_{0}(t+T)$ of an Ising chain:
\begin{equation}\label{eq_tfih}
H_{0}(t)=\begin{cases}
		g \sum_{j=1}^{L} \sigma_{j}^{z} & \text{for ~~$0 < t < T/2$}, \\
		 J \sum_{j=1}^{L-1} \sigma_{j}^{x} \sigma_{j+1}^{x} & \text{for ~~$T/2< t < T$},
	\end{cases}
\end{equation}
where the $\sigma$'s are Pauli matrices, $g$ is the transverse field, and $J$ is the coupling strength between nearest-neighbour spins. 
\tb{This model is integrable and can be mapped to a Kitaev chain of spinless fermions (with a driving protocol different from that we have explored in the previous sections) through Jordan-Wigner transformations~\cite{lieb61}. The periodically driven Kitaev chain can host Floquet Majorana edge-modes.} For the particular choice of the parameters $J=0.5$ and $g=0.6$, the first Floquet Majorana $\pi$-mode at each edge of the chain is generated at $\omega=2.2$, as the Floquet quasienergy gap becomes $\pm \pi/T$ at $\omega=2.2$ (see the analytical calculation presented in Appendix~\ref{app_igap}). Thus, the Floquet Majorana modes exist only for $\omega < 2.2$ for this particular choice of the parameters.\\

The integrability of the model can be broken by adding a longitudinal field term in the Hamiltonian $H_{0}(t)$. Following the driving protocol as in Ref.~\cite{mi22}, we include a site-dependent, random longitudinal field ($h_{j}$) term (with 
$\delta$-function kicks) in the Hamiltonian $H_{0}(t)$ to test the robustness of the Floquet Majorana modes against the integrability breaking interaction. Thus, the time-dependence of the Hamiltonian is given by
\begin{equation}\label{eq_ilf}
H(t)=H_{0}(t) + \frac{1}{2}\sum_{j=1}^{L} h_{j} \sigma_{j}^{x} \sum_{n=-\infty}^{\infty}\delta(t-nT),
\end{equation}
where the longitudinal field $h_{j}$ for the $j$-th site is a random variable which can assume any value lying in the range $[-\pi,\pi]$, 
for all $j = 1, 2, \cdots, L$. Thus, the Floquet operator is given by
(see also Eq.~(1) of Ref.~\cite{mi22}),
\begin{widetext}
\begin{equation}\label{eq_Uif}
U_{F} = \exp(-\frac{i}{2} \sum_{j=1}^{L} h_{j} \sigma_{j}^{x}) \exp(-i \frac{J T}{2} \sum_{j=1}^{L-1} \sigma_{j}^{x} \sigma_{j+1}^{x}) \exp(-i \frac{g T}{2}\sum_{j=1}^{L}\sigma_{j}^{z}) .
\end{equation}
\end{widetext}

We find that at $\omega=2.0$ (i.e., $T=2\pi/\omega=\pi$), the DEE ($S_{D}$) is equal to $\ln(2)$ in the integrable limit (i.e., $h_{j}=0$) with parameters $J=0.5$ and $g=0.6$ (see the plot for $h_{j}=0$ in Fig.~\ref{fig_hx}), as there is one Floquet Majorana $\pi$-mode at each edge, namely, there are two Floquet Majorana $\pi$-modes in total at $\omega=2.0$.\\

Using the exact diagonalization scheme for the non-integrable case and open boundary conditions, we find that in the presence of the random, site-dependent longitudinal field, i.e., when $h_{j} \neq 0$ (see Fig.~\ref{fig_hx}), the DEE for system size $L=4L_{D}=12$ remains close to $\ln(2)$ at $\omega=2.0$. This result indicates the robustness of the Floquet Majorana modes against an integrability breaking interaction. The deviation of the DEE from perfectly integer-quantized values at $\omega=2.0$ in the presence of the longitudinal field is due to the finite size $L$ of the system, as elaborated in Sec.~\ref{sec_drive}. The DEE deviates from the integer-quantized values more and more when the frequency is decreased below $\omega=2.0$. It is also important to mention here that due to numerical limitations, the maximum size of the system for which the DEE in the non-integrable case and open boundary conditions can be computed using the exact diagonalization scheme is given by $L=4L_{D}=12$. Thus, numerical limitations prohibit us from obtaining perfectly integer-quantized value of the DEE for the drive frequencies $\omega \leq 2.0$ in the presence of the integrability breaking interaction. Therefore, one cannot conclusively comment on the fate of the Floquet Majorana modes in the presence of arbitrary integrability breaking perturbations, and the results presented in Fig.~\ref{fig_hx} indicate
their robustness at least for some values of the parameters.
 
\section{Conclusions}
\label{sec_conclusion}

In this paper, we have shown that for the periodic driving protocol of the 
chemical potential with $\delta$-pulses, the DEE, calculated in the ground state of the Floquet 
Hamiltonian, is integer-quantized, with the integer being equal to the number of 
dynamically generated Floquet Majorana edge modes localized at each edge of the chain. Thus, it can be inferred that 
similar to the static situation, the DEE can act as a marker of Majorana edge modes even for a periodically driven Kitaev chain.
At low frequencies, there is an apparent discrepancy and the value of $S_{D}/\ln(2)$
differs significantly from the number of Majorana end modes at each edge of the chain. However, this is an 
artefact of the finite size of the system, and we have established 
that the DEE saturates to an integer-quantized value at large system size $L$, even 
at low drive frequencies ($\omega$ of the order of the hopping amplitude $\gamma$).
Thus, in a periodically driven chain with open boundary conditions, the DEE 
counts the number of edge Majorana modes correctly and therefore plays a role similar to that played by the winding number derived from the Floquet Hamiltonian in the corresponding system with periodic boundary conditions. Similar results can also be obtained for other periodic driving protocols (for instance, square pulse and sinusoidal modulation) of the chemical potential. Further, by probing the DEE, we have shown the robustness of the Floquet Majorana edge modes (zero modes and $\pi$-modes) against weak spatial disorder and temporal noise.

If the amplitude of the nearest-neighbour hopping in the Kitaev chain is periodically modulated (see the protocol in Eq.~\eqref{eq_sym}), such that the Floquet Hamiltonian remains in the BDI symmetry class, then 
``anomalous" edge modes (with Floquet quasienergies not equal to zero or $\pi$) can be
dynamically generated. Although these anomalous edge modes do not have a topological
origin and are not associated with a winding number, we find that the DEE is 
able to detect the existence of the anomalous edge modes 
for certain ranges of driving frequencies. Furthermore, we illustrate that the anomalous edge modes are not robust against weak spatial disorder, and they disappear in that situation while the zero-energy Majorana modes survive.\\

The dynamical generation of anomalous edge modes is also possible with the periodic modulation of the phase of the complex hopping parameter (see the driving protocol in Eq.~\eqref{eq_brk}, where the system belongs to Floquet D symmetry class). However, for this driving protocol, the DEE fails to detect them properly. Further, we have checked that one cannot infer conclusively
the effects of disorder on both zero-energy and anomalous modes in the presence of spatial disorder. 
In the symmetry class D, time-reversal symmetry is broken and the Majorana edge-modes are not protected from hybridization. Due to the hybridization of the edge-modes with each other, we find that the value of $S_{D}/\ln(2)$ is different from the total number of Majorana modes and anomalous modes localized at each end of the chain for the symmetry class D. Therefore, if the edge-modes are not protected by any one (or more than one) of time-reversal symmetry, particle-hole symmetry and sub-lattice symmetry, then the value of $S_{D}/\ln(2)$ may differ from the number of edge-modes localized at each edge. However, it might be an interesting future work to investigate whether the DEE is able to detect the edge-modes for other symmetry classes, especially when any one of the above mentioned symmetries is not present. It might also be a possible future work to compute the DEE for systems with half-integer spins in the symmetry class CII and DIII.\\ 

We have also explored the behavior of the DEE in the presence of an integrability breaking perturbation in a kicked Ising chain, which has recently been realized experimentally~\cite{mi22}. However, due to numerical limitations in calculating the DEE for a sufficiently large system size in the non-integrable situation and open boundary conditions using the exact diagonalization scheme, we are unable to 
verify the robustness of the Floquet Majorana modes against an integrability breaking perturbation by probing the DEE.

We conclude by noting that recently the topological entanglement entropy has been measured experimentally for a two-dimensional toric code model \cite{satzinger21}; the quantity has been measured experimentally using simulated anyon interferometry and extracting the braiding statistics of emergent excitation~\cite{satzinger21}. On the other hand, in the present work, we have computed the DEE for one-dimensional Kitaev model where anyons do not exist. We are also not aware of an equivalent of the DEE in two-dimensional situations. This experiment however provides a motivation for an experimental measurement of the DEE for one-dimensional topological systems and a possible generalization of the DEE for two-dimensional systems.

Finally, we note that it would be interesting to investigate out-of-equilibrium behavior of the DEE in Kitaev chains with more complicated couplings, like next-nearest-neighbor hopping and superconducting pairing~\cite{trs_sen13, bsourav18} and multi-critical situations~\cite{Mukherjee_2010}.

\begin{acknowledgments}
S.M. acknowledges financial support from PMRF fellowship, MHRD, India. D.S. thanks SERB, India for funding through Project No. JBR/2020/000043. A.D. acknowledges support from SPARC program, MHRD, India and SERB, DST, New Delhi, India. We acknowledge Souvik Bandyopadhyay and Sourav Bhattacharjee for comments. AD acknowledges Marcello Dalmonte and Vittorio Vitale for discussion and the associateship programme, Abdus Salam ICTP, Italy.
\end{acknowledgments}

\appendix

\section{Winding number and topological properties of short-range Kitaev chain}\label{sec_topology}
The Hamiltonian $H$ in Eq.~\eqref{eqn_H} of a short-range Kitaev chain with periodic boundary conditions can be written in terms the fermionic creation and annihilation operators in momentum space as
\begin{equation}\label{eqn_H_momentum}
H = \sum_{k} \begin{pmatrix} c^{\dagger}_{k} & c_{-k} \end{pmatrix} 
H_{k} \begin{pmatrix}
c_{k} \\
c^{\dagger}_{-k} \end{pmatrix}, \end{equation}
where $k$ lies in the range $[-\pi,\pi]$, and the Hamiltonian $H_{k}$ is given by
\begin{equation}\label{eqn_H_k}
H_{k}=(-\gamma \cos k -\mu) ~\tau_{z} + \Delta \sin k ~\tau_{y}, \end{equation}
where $\tau_{y}$ and $\tau_{z}$ are Pauli matrices.
In the ground state of the Hamiltonian $H$ in Eq.~\eqref{eqn_H}, the winding number~\cite{dsen13} is defined as
\begin{eqnarray}\label{eq_winding}
w &=& \frac{1}{2 \pi} \int_{-\pi}^{\pi} dk \frac{d \phi_{k}}{dk}, \\
\phi_{k} &=& \tan^{-1} \left(\frac{\Delta \sin k}{\gamma \cos k +\mu} \right).
\end{eqnarray}

Setting the hopping parameter $\gamma=1$, it is straightforward to check that the following phases exist in the ground state of a short-range Kitaev chain (in static situation).
\begin{enumerate}

\item Topologically non-trivial phase ($-1<\mu<1$): This phase consists of phase I ($\Delta>0$) and phase II ($\Delta<0$). The winding numbers in phases I and phases II are given by $w=+1$ and $w=-1$, respectively~\cite{dsen13}. For a chain with open boundary condition, these topologically non-trivial phases are characterized by the existence of a zero-energy Majorana mode localized at each edge of the chain~\cite{dsen13}.

\item Topologically trivial phase ($|\mu| >1$): The winding number is zero ($w=0$) in this phase. For a chain with open boundary condition, Majorana edge modes do not appear in this phase.
\end{enumerate}

\section{Methods used for calculation of the DEE from Floquet Hamiltonian}\label{app_method}

Each fermionic creation operator $c_{n}^{\dagger}$ (or annihilation operator $c_{n}$) can be written as a linear combination of two Hermitian Majorana operators ($a_{2n-1}$ and $a_{2n}$) as
\begin{subequations}\label{eq_majorana}
\begin{equation}
c_{n}=\frac{1}{2} (a_{2n-1} - i a_{2n}),
\end{equation}
\begin{equation}
c_{n}^{\dagger}=\frac{1}{2} (a_{2n-1} + i a_{2n}).
\end{equation}
\end{subequations}
The Hamiltonian $H(t)$ of a Kitaev chain with generic time-dependent parameters can be written as,
\begin{equation}\label{eq_H_majorana}
H(t)=i \sum_{m=1}^{2L} \sum_{n=1}^{2L} a_{m} M_{mn} (t) a_{n},
\end{equation}
where $M(t)$ is a $2L \times 2L$ real, antisymmetric matrix (which follows from the fact that Majorana operators satisfy anticommutation relations: $\{ a_{m},a_{n} \}=2\delta_{mn}$). The $2L \times 1$ column matrix of the Majorana operators $a(t)=\begin{pmatrix} a_{1}(t) & a_{2}(t) & .& . & . & a_{2L}(t) \end{pmatrix} ^{T}$ in Heisenberg picture can be written as,
\begin{equation}\label{eq_a_U}
a(t)=\mathbb{T} \exp(4 \int_{0}^{t} M(t^{\prime}) dt^{\prime}) a (0) = U(t) a (0).
\end{equation}
Thus, the stroboscopic time-evolution operator (Floquet operator) is given by,
\begin{equation}\label{eq_Ut}
U_{F}=U(T)=\mathbb{T} \exp(4 \int_{0}^{T} M(t^{\prime}) dt^{\prime}).
\end{equation}
As $U_{F}$ is unitary matrix, each of its eigenvalues $\lambda$ satisfies the property $|\lambda|=1$. Thus, $\lambda$ must be phases. Further, the matrix $U_{F}$ is real, which ensures that the eigenvalues of $U_{F}$ must appear in complex conjugate pairs, namely, $e^{i\theta}$ and $e^{-i\theta}$.

For a subsystem $X$ with $L_{X}$ fermionic sites, an element of the $2L_{X} \times 2L_{X}$ correlation matrix $A_{X}$, calculated in the ground state of the Floquet Hamiltonian, can be written as
\begin{equation}\label{eq_corr}
(A_{X})_{mn}=\langle a_{m}a_{n} \rangle.
\end{equation}
The von Neumann entropy $S_{X}$ of the subsystem $X$ is obtained from the eigenvalues $\alpha_{j}$ of the correlation matrix $A_{X}$ as
\begin{equation}\label{eq_entropy}
S_{X}= - \sum_{j=1}^{2L_{X}} \frac{\alpha_{j}}{2} \ln(\frac{\alpha_{j}}{2}). 
\end{equation}
$S_{X}$ for $X=A,B,A\cap B$ and $A\cup B$ are calculated in the ground state of Floquet Hamiltonian and the DEE is then computed using Eq.~\eqref{eqn_SD}.

\section{Comparison of the DEE with the dynamical winding number for periodically driven Kitaev chain}
\label{sec_winding}

\begin{figure*}
\centering
\subfigure[]{
\includegraphics[width=0.45\textwidth]{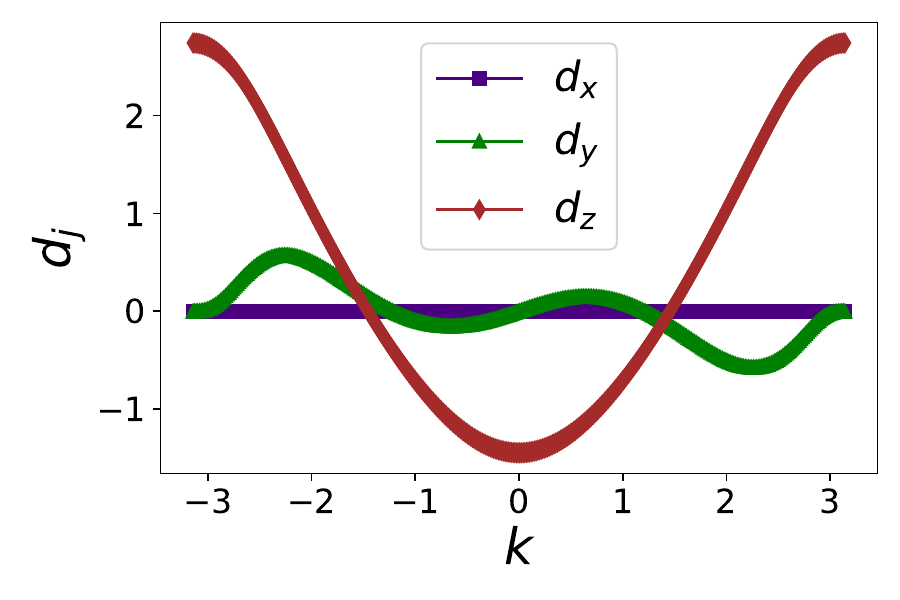}
\label{fig_kd_delta}}
\hspace{5mm}	
\centering
\subfigure[]{
\includegraphics[width=0.45\textwidth]{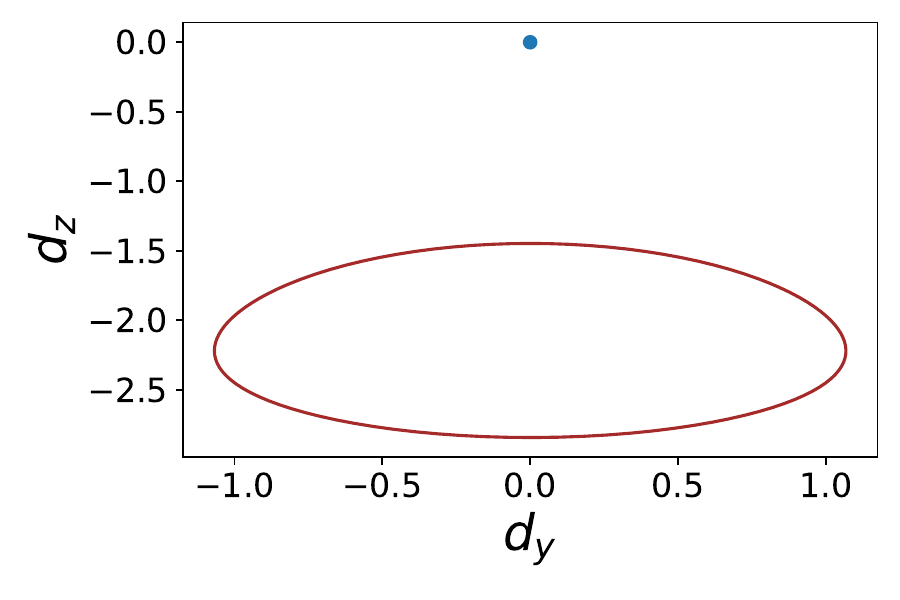}
\label{fig_w_zero}}	
\centering
\subfigure[]{
\includegraphics[width=0.45\textwidth]{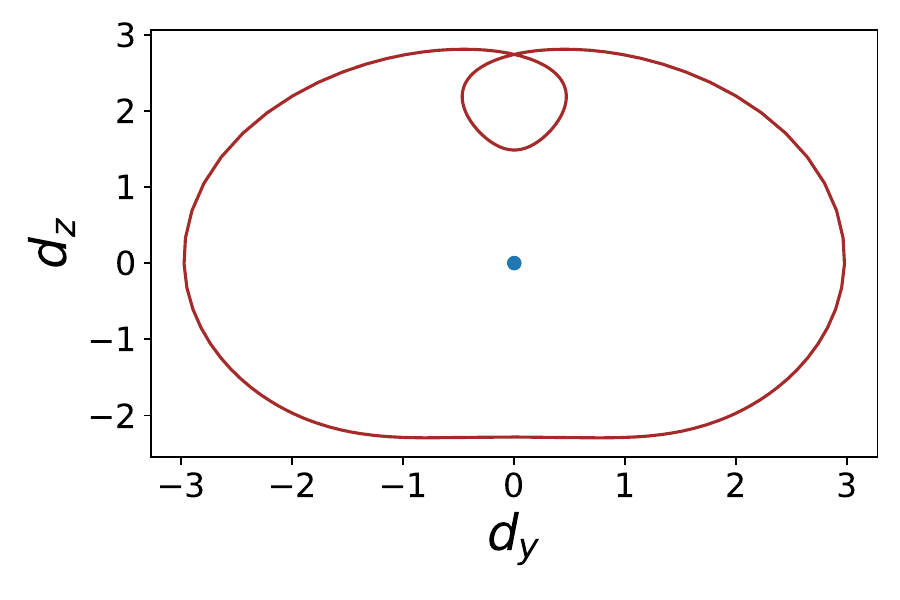}
\label{fig_w_one}}
\hspace{5mm}	
\centering
\subfigure[]{
\includegraphics[width=0.45\textwidth]{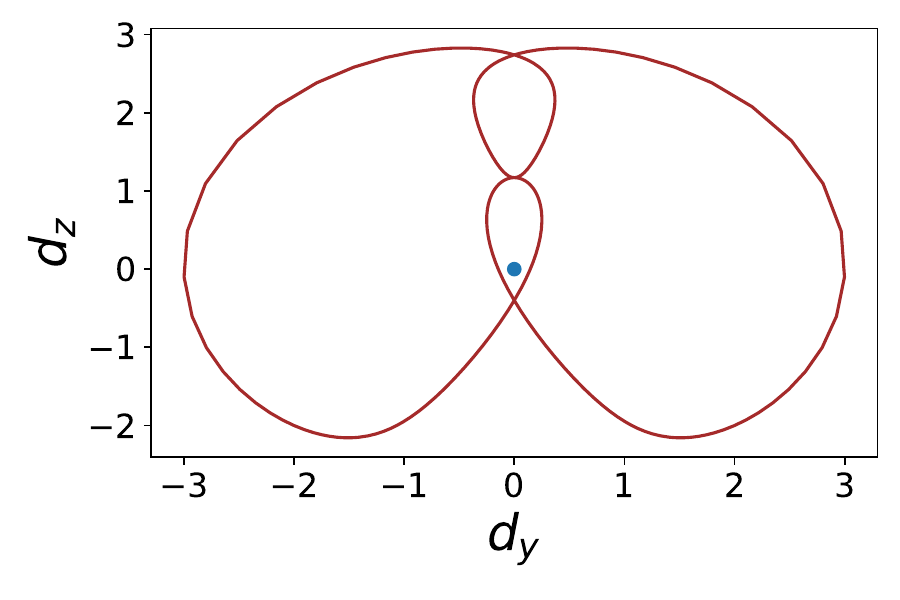}
\label{fig_w_two}}
\centering
\caption{(a) $d_{x}$, $d_{y}$ and $d_{z}$ as functions of $k$ for periodic 
driving of the chemical potential with a $\delta$-pulse having $\mu_{0}=2.5$,
$\mu_{1}=0.2$, and $\omega=6.0$. We find that $d_{x}=0$ for all values 
of $k$. (b) $d_{z}$ as a function of $d_{y}$ for $\omega=18.0$. The dynamical (Floquet) winding number is
zero. (c) $d_{z}$ as a function of $d_{y}$ for $\omega=10.0$. The winding number 
is $1$. (d) $d_{z}$ as a function of $d_{y}$ for $\omega=4.0$. The 
winding number is $2$.} \label{fig_w_delta} \end{figure*}
In this section, we compare the results inferred from the behavior of the DEE with 
the dynamical winding number for driving protocol discussed in Sec.~\ref{sec_drive}.

For a periodically driven Kitaev chain with periodic boundary conditions, the winding 
number may be calculated from the stroboscopic time-evolution operator (i.e., Floquet 
operator)
\begin{equation}\label{eq_ukf}
U_{F} (k) = \mathbb{T} \exp(-i \int_{0}^{T} H_{k} (t) dt) = \exp(-i h_{k}^{F} T),
\end{equation}
for momentum $k$ lying in the range $[-\pi,\pi]$, 
where $h_{k}^{F}$ and $H_{k} (t)$ are, 
respectively, the Floquet Hamiltonian and the instantaneous Hamiltonian for the 
mode with momentum $k$. Referring to Eq.~\eqref{eqn_H_k}, for a generic
time-dependent chemical potential, we get
\begin{equation}\label{eq_hkt}
H_{k} (t) = (-\gamma \cos k -\mu (t)) ~\tau_{z} + \Delta \sin k ~\tau_{y}.
\end{equation}
On the other hand, for a periodic driving, the general form of the Floquet Hamiltonian 
$h_{k}^{F}$ can be written as
\begin{equation}\label{eq_hkf_d0}
h_{k}^{F} = d_{0} (k) \mathbb{1} + d_{x} (k) \tau_{x} + d_{y} (k) \tau_{y} + 
d_{z} (k) \tau_{z}, \end{equation}
where $\tau_{x}$, $\tau_{y}$, $\tau_{z}$ are Pauli matrices and $\mathbb{1}$ is $2 \times 2$ identity matrix.
Since $U_{F} (k)$ is a $SU(2)$ matrix (due to the fact that $H_{k}(t)$ can be written as the sum of Pauli matrices with appropriate coefficients), we have $d_{0} (k) =0$ for all values of $k$. Thus, the Floquet Hamiltonian 
$h_{k}^{F}$ ~\cite{dsen13, saha17} reduces to the form
\begin{equation}\label{eq_hkf}
h_{k}^{F} = d_{x} (k) \tau_{x} + d_{y} (k) \tau_{y} + 
d_{z} (k) \tau_{z} . \end{equation}
The coefficients $d_{x}$, $d_{y}$ and 
$d_{z}$ assume different forms for different driving protocols.

If the chemical potential is periodically modulated with $\delta$-pulses 
(Eq.~\eqref{eq_delta}), the symmetrized Floquet operator $U_{F} (k)$, for $k$
lying in the range $[-\pi, \pi]$, is given by
\begin{equation}\label{eq_ufk_delta}
U_{F} (k) = e^{i \frac{\mu_{1}}{2}\tau_{z}} e^{-i T \left[ (-\gamma \cos k 
-\mu_{0}) ~\tau_{z} + \Delta \sin k ~\tau_{y} \right]} e^{i 
\frac{\mu_{1}}{2}\tau_{z}}. 
\end{equation}
Since we have chosen the driving to satisfy $\mu(t)=\mu(T-t)$ and $\tau_x H_k (t) \tau_x =
- H_k (T-t)$, we see that $\tau_x U_F (k) \tau_x = [U_F (k)]^{\dagger}$) for all 
$k$. Using $U_{F}(k)=\exp(-ih_{k}^{F}T)$ and Eq.~\eqref{eq_hkf}, we find that $d_x (k) = 0$; we can see in Fig.~\ref{fig_kd_delta} that this is true.
We then arrive at a simplified form of the Floquet Hamiltonian $h_{k}^{F}$ 
\begin{equation}\label{eq_hkf_delta}
h_{k}^{F} = d_{y} (k) \tau_{y} + d_{z} (k) \tau_{z}. \end{equation}
This particular form of $h_{k}^{F}$ enables us to 
define a dynamical winding number~\cite{dsen13} in the following way: $d_{z} (k)$ 
is plotted as a function of $d_{y} (k)$ (see Figs.~\ref{fig_w_zero}, \ref{fig_w_one} 
and \ref{fig_w_two}) and the number of times the curve winds around the origin (located 
at $d_{y}=0$, $d_{z}=0$) is counted. This gives the absolute value of the Floquet winding number ($|w|$).

The Floquet winding number can also be obtained using the following method. For a generic time-dependent $\mu(t)$, the Hamiltonian $H_{k}$ for $k$ lying in the
range $[0, \pi]$ is given by
\begin{equation}\label{eq_muH}
H_{k} (t) = (-2 \gamma \cos(k) - 2\mu(t)) \tau_{z} + 2 \Delta \sin(k) \tau_{y}.
\end{equation}
For the periodic modulation of chemical potential $\mu(t)$ with a sequence of $\delta$-pulses as in Eq.~\eqref{eq_delta} of the main text, the Floquet operator $U_{F}(k)$ is given by
\begin{equation}\label{eq_ufd}
U_{F} (k) = e^{i \mu_{1} \tau_{z}} e^{-2 i T \left[ (-\gamma \cos k 
-\mu_{0}) ~\tau_{z} + \Delta \sin k ~\tau_{y} \right]} e^{i \mu_{1} \tau_{z}}.
\end{equation}
For $k=0$ and $k=\pi$, the Floquet operators ($U_{F}(k)$) are:
\begin{subequations}
\begin{equation}\label{eq_k0d}
U_{F} (k=0)= \exp(i(2 \mu_{1} + 2 T (\gamma +\mu_{0}))\tau_{z}), 
\end{equation}
\begin{equation}\label{eq_kpd}
U_{F} (k=\pi)= \exp(i(2 \mu_{1} + 2 T (\mu_{0} - \gamma))\tau_{z}).
\end{equation}
\end{subequations}
The Floquet operators $U_{F} (k=0)$ and $U_{F} (k=\pi)$ can also be written as, $U_{F} (k=0)=\exp(i\pi b_{0}\tau_{z})$ and $U_{F} (k=\pi)=\exp(i\pi b_{\pi}\tau_{z})$. Here, $b_{0}$ and $b_{\pi}$ are given by the following equations:
\begin{subequations}
\begin{equation}\label{eq_b0}
b_{0}= \frac{2\mu_{1}}{\pi}+ \frac{4}{\omega} (\gamma +\mu_{0}), 
\end{equation}
\begin{equation}\label{eq_bpi}
b_{\pi}= \frac{2\mu_{1}}{\pi}+ \frac{4}{\omega} (\mu_{0}-\gamma).
\end{equation}
\end{subequations}
The number of Floquet Majorana modes at each end of the chain with quasienergies zero (or eigenvalue $+1$ of the Floquet operator) and $\pm \pi/T$ (or eigenvalue $-1$ of the Floquet operator) are given by, $p(0)=|n_{e}^{>}-n_{e}^{<}|$ and $p(\pi/T)=|n_{o}^{>}-n_{o}^{<}|$ (as shown in Ref.~\cite{dsen13}), where $n_{e}^{>}$ ($n_{o}^{>}$) is the number of even (odd) integers lying between $b_{0}$ and $b_{\pi}$, and greater than $2\mu_{1}/\pi$. On the other hand, $n_{e}^{<}$ ($n_{o}^{<}$) is the number of even (odd) integers lying between $b_{0}$ and $b_{\pi}$, and less than $2\mu_{1}/\pi$. The winding number~\cite{dsen13} is determined by, $|w|=|n_{e}^{>}-n_{e}^{<}+ n_{o}^{>}-n_{o}^{<}|$. Therefore, in general, we have $|w| \leq p(0) + p(\pi/T)$. If $p=p(0)+p(\pi/T)$ is the total number of Floquet Majorana modes at each end of the chain, then, in general, we have $|w| \leq p$~\cite{dsen13, russomanno17}. However, if we choose $\mu_{0}>\gamma$ (where $\gamma>0$ and $\mu_{0}>0$), then for $0< 2\mu_{1}/\pi<1$, $n_{o}^{<}=n_{e}^{<}=0$. Thus, for $0<\mu_{1}<\pi/2$, we have $p(0)=n_{e}^{>}$, $p(\pi/T)=n_{o}^{>}$ and $|w|=n_{e}^{>} + n_{o}^{>}$. Therefore, $|w|=p(0)+p(\pi/T)=p$, if $0< \mu_{1}<\pi/2$. In this situation, the winding number counts the Floquet Majorana modes (with
quasienergies both zero and $\pm \pi/T$) at each end of the chain correctly. 
Thus, the winding number is able to characterize the topology of 
the Kitaev chain periodically driven with a $\delta$-pulse.

On the other hand, the DEE, which can assume any real value, turns out to be an integer multiple of $\ln(2)$, namely, $S_{D}= p\ln(2)$ (see Fig.~\ref{fig_delta}); here, the integer $p$ is the number of Floquet Majorana modes localized at each edge (with quasienergies 
both zero and $\pm \pi/T$). 
Comparing with the Floquet winding number $w$, we find that $|w|\ln(2) \leq S_{D}$, in general. However, if we choose the parameters in the following way, $\gamma>0$, $\mu_{0}>0$, $\mu_{0}>\gamma$ and $0<\mu_{1}<\pi/2$, then we obtain $|w|\ln(2)=S_{D}$. This
establishes the equivalence of the DEE with the winding number in detecting the
Majorana edge modes for the periodic modulation with a $\delta$-pulse.

\section{Floquet quasienergy gap in a periodically driven chain}\label{app_gap}

\subsection{For periodic driving of hopping parameter}\label{app_hgap}
The Floquet operator for the mode with momentum $k$ lying in the range $[0,\pi]$ 
is given by
\begin{equation}\label{eq_ukfh}
U_{F} (k) = \mathbb{T} \exp(-i \int_{0}^{T} H_{k} (t) dt) = \exp(-i h_{k}^{F} T),
\end{equation}
where $h_{k}^{F}$ is the corresponding Floquet Hamiltonian. For a time-dependent hopping 
amplitude $\gamma(t)$, $H_{k}(t)$ has the form
\begin{equation}\label{eq_hkh}
H_{k} (t) = (-2 \gamma (t) \cos k -2 \mu) ~\tau_{z} + 2 \Delta \sin k ~\tau_{y},
\end{equation}
where $\gamma(t)=\gamma_{0}(1+a\cos(\omega t))$ (see Eq.~\eqref{eq_sym}). Using Eq.~\eqref{eq_ukfh}, we obtain the following equation for $k=0$:
\begin{equation}\label{eq_k0}
U_{F} (k=0)= e^{2 i T (\gamma_{0}+\mu) \tau_{z}}.
\end{equation}
For $\mu=0$, the condition $U_{F}(k=0)=\mathbb{1}$ is satisfied at the drive frequency $\omega=2\gamma_{0}/n$, where $n$ is an integer. At these frequencies,
the Floquet quasienergy gap closes. Thus, the generation of zero-energy Floquet Majorana modes occurs at these specific frequencies for the driving protocol given in Eq.~\eqref{eq_sym} in a chain with open boundary condition. For the periodic driving of the chemical potential, an analytical derivation of the frequencies at which the Floquet quasienergy gap closes can also be done by proceeding in the same way as discussed in this section (See Ref.~\cite{dsen13}).\\

\subsection{For the driving protocol in Eq.~\eqref{eq_tfih} for an Ising chain}\label{app_igap}
Using the Jordan-Wigner transformation~\cite{lieb61} and periodic boundary conditions for the driving protocol in Eq.~\eqref{eq_tfih} for an Ising chain, the Floquet evolution operator $U_{F} (k)$ for the mode with momentum $k$ lying
in the range $[0,\pi]$ can be written as
\begin{equation}\label{eq_utf}
U_{F} (k) = e^{-i \frac{T}{2} \left( 2J \cos(k) \tau_{z} - 2J \sin(k) \tau_{y} \right)} e^{-i \frac{T}{2} (2g \tau_{z})},
\end{equation}
where $\tau_{y}$ and $\tau_{z}$ are Pauli matrices. For the mode with momentum $k=0$, we have,
\begin{equation}\label{eq_k0i}
U_{F} (k=0)= e^{-i T (J+g) \tau_{z}}.
\end{equation}
The Floquet quasienergy gap becomes zero and $\pm \pi/T$ if $U_{F}(k=0)=\mathbb{1}$ and $U_{F}(k=0)=-\mathbb{1}$, respectively. From these two conditions, we see that the Floquet quasienergy gap becomes zero and $\pm \pi/T$ at the drive frequencies $\omega=(J+g)/n$ and $\omega=(2J+2g)/(2n+1)$, respectively, where $n$ is an integer. For the parameters $J=0.5$ and $g=0.6$ chosen for the numerical computations in Sec.~\ref{sec_hx}, it can be seen that the maximum values of the drive frequencies at which the Floquet quasienergy gap becomes zero and $\pm \pi/T$ are $\omega=1.1$ and $\omega=2.2$, respectively.

\bibliography{reference_sd_periodic}

\end{document}